\documentclass{article}
\usepackage{graphicx}

\begin{document}

\title{Some remarks on black hole thermodynamics}
\date{February 4, 2011}
\author{Raymond Y. Chiao\\Emeritus Professor\\University of California at Merced\\P.O. Box 2039\\Merced,CA 95344\\rchiao@ucmerced.edu}
\maketitle

\begin{figure}[tbp]
\includegraphics[width=5in]{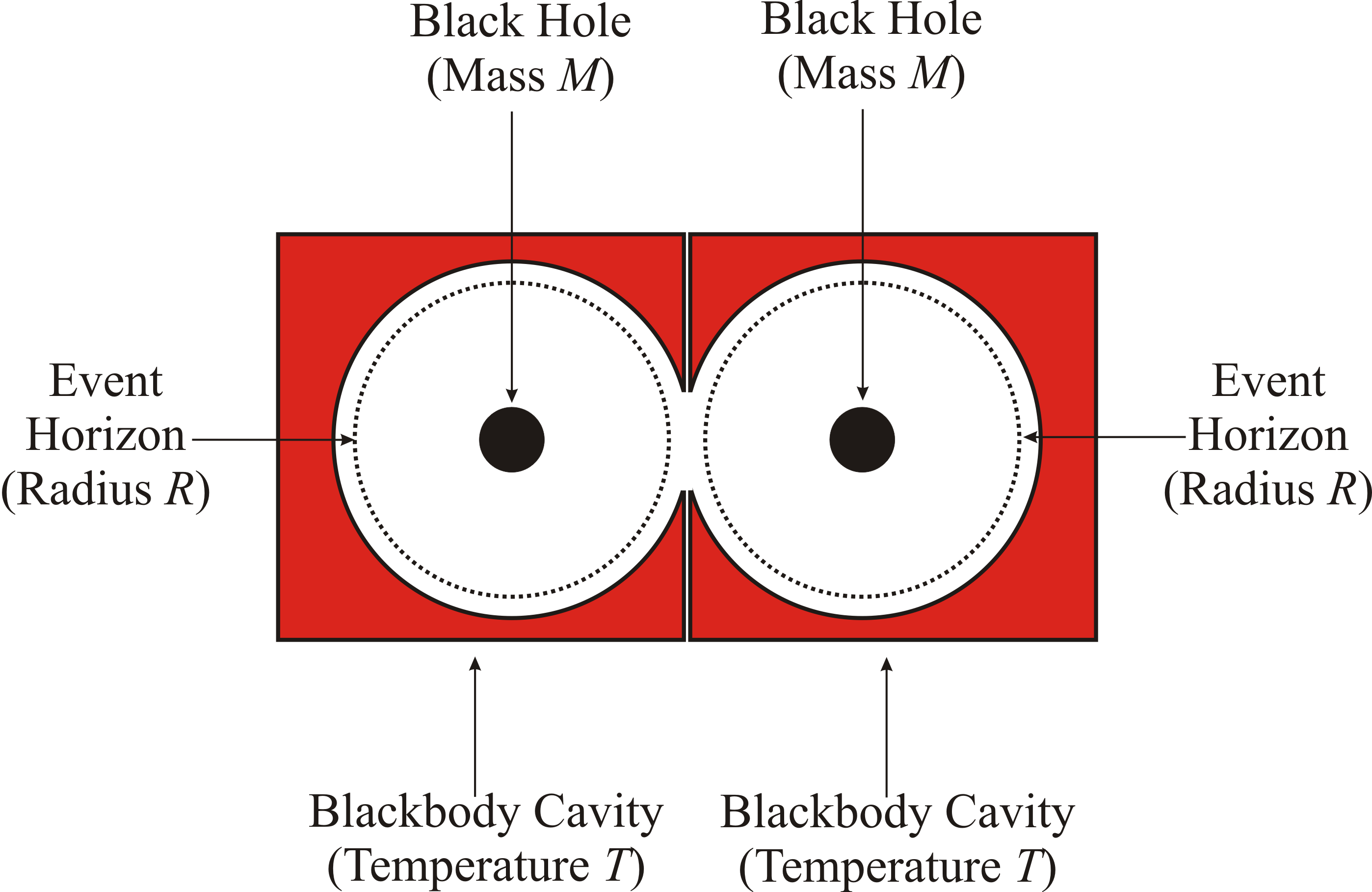}
\caption{Symmetrical arrangement of two blackbody cavities with two
identical black holes with mass $M$ at their centers. The two event horizons
have the same radius $R$, which just fits inside the spherical blackbody
cavities. These cavities are coupled to each other via two identical
apertures, so that the two cavities come into thermodynamic equilibrium at
the same temperature $T$. The walls of the two cavities are composed of
thermally insulating material. This symmetric configuration is thermally
unstable (see text).}
\end{figure}

The purpose of this White Paper is to make some remarks on
foundational questions concerning two thermodynamic
\textquotedblleft paradoxes\textquotedblright\ in black hole physics. The
first \textquotedblleft paradox\textquotedblright\ involves involves the
thermal instability of two identical black holes placed at the centers of two
identical blackbody cavities (see Figure 1), and the second
\textquotedblleft paradox\textquotedblright\ involves the apparent violation
of the second law by a classical black hole in thermodynamic
equilibrium with a heat bath at a finite temperature $T$, which can absorb
incoming thermal radiation, but cannot emit outgoing thermal radiation.

In the first apparent thermodynamic \textquotedblleft
paradox,\textquotedblright\ let us consider the following simple thought
experiment: Two spherical blackbody cavities, which are facing with other in close
proximity, contain at their centers two identical black holes with the same mass $M$%
, as illustrated in Figure 1. The event horizons of the two black holes have a
radius $R$ which is just slightly smaller than the radius of the spherical
cavities. The two blackbody cavities are coupled to each other by means of two
small, matching apertures of the same size and shape, so that they can come
into thermodynamic equilibrium at the same temperature $T$ with respect to
each other.

Is the resulting thermodynamic equilibrium between the two identical black holes
emitting blackbody radiation at the same Hawking temperature $T$, which are
immersed in heat baths inside the two blackbody cavities at the same blackbody
temperature $T$, stable or unstable?

In order to answer this question, let us begin with the Hawking temperature $%
T$ of a zero-charge, zero-angular-momentum black hole, which is given by \cite%
{Hawking} \cite{Unruh}%
\begin{equation}
k_{B}T=\frac{1}{2\pi }\frac{\hbar g}{c}  \label{Hawking temperature}
\end{equation}%
where $k_{B}$ is Boltzmann's constant, $\hbar $ is the reduced Planck's
constant, $c$ is the speed of light, and%
\begin{equation}
g=\frac{GM}{R^{2}}  \label{surface gravity of black hole}
\end{equation}%
is the \textquotedblleft surface gravity\textquotedblright\ of the black
hole, i.e., the acceleration due to gravity of a test mass located at the
Schwarzschild radius $R$, i.e., at the event horizon of the black hole,%
\begin{equation}
R=\frac{2GM}{c^{2}}  \label{Schwartzschild radius}
\end{equation}%
where $G$ is Newton's constant, and $M$ is the mass of the black hole.
Substituting (\ref{Schwartzschild radius}) into (\ref{surface gravity of black hole}%
), one finds that%
\begin{equation}
g=\frac{c^{4}}{4GM}  \label{surface gravity of black hole in terms of its mass}
\end{equation}%
Substituting (\ref{surface gravity of black hole in terms of its mass})\ into (\ref%
{Hawking temperature}), one finds that%
\begin{equation}
k_{B}T=\frac{1}{8\pi }\frac{\hbar c^{3}}{GM}
\label{T inversely proportional to M}
\end{equation}%
This implies that%
\begin{equation}
T\propto \frac{1}{M}
\end{equation}

Therefore if the mass of one of the black holes, say the one in the left blackbody cavity,
were to \textit{decrease} due to a fluctuation \cite{fluctuation}, its
temperature would \textit{increase}. However, such a temperature \textit{%
increase} would result in the left blackbody cavity becoming \textit{hotter}. This
would imply an \textit{increase} in the emission of blackbody photons from the left
cavity into the right blackbody cavity. The extra photons flowing from the left
cavity to the right cavity would be swallowed up by the right black hole.

Hence, by energy conservation, the \textit{increased} rate of emission of
energy by the left black hole due a fluctuation implies that its mass must \textit{%
decrease}, and that the mass of the right black hole must \textit{increase,} in a
compensatory manner. But an \textit{increase} in the mass of the right black hole
would mean a $decrease$ of its temperature, which would imply a further 
\textit{decrease} in the mass of the left black hole, which would mean a further 
\textit{increase} of its temperature, etc. This would lead to a \textit{%
thermal runaway} phenomenon, i.e., an \textit{instability} in which the
system runs away from thermodynamic equilibrium. To a distant observer, the
two blackbody cavities, which were originally at the same temperature, would
appear to spontaneously develop a larger and larger difference in their
temperatures over time.

This would seem to violate one of the common statements of the second law of
thermodynamics, for example, the following textbook statement \cite[p.619]%
{Halliday & Resnick}:

\begin{quote}
\textquotedblleft The first law of thermodynamics states that energy is
conserved. However, we can think of many thermodynamic processes which
conserve energy but which actually never occur. For example, when a hot body
and a cold body are put into contact, it simply does not happen that the hot
body gets hotter and the cold body colder.\textquotedblright
\end{quote}

One can rephrase the last statement more precisely as follows [using the
author's words]:

\begin{quote}
\textquotedblleft It is impossible for two bodies in thermal contact with
each other, which are initially in thermodynamic equilibrium at the same
temperature, to spontaneously depart from this equilibrium, such that one
body steadily increases in temperature and the other steadily decreases in
temperature over time.\textquotedblright
\end{quote}

The intuition behind these putative statements of the second law is that if
a movie were to be made of the behavior of the two thermally connected
bodies, which could be two \textquotedblleft black boxes\textquotedblright\
with thermometers sticking out of them in order to measure their
temperatures, and with a thermally conducting strap connecting them together
in order to establish thermal equilibrium, the natural \textquotedblleft
arrow of time\textquotedblright\ for the movie would be for the two black
boxes to come into equilibrium at the same temperature over time. It would
seem to be unnatural for the \textquotedblleft arrow of
time\textquotedblright\ to point in the opposite direction, i.e., for the
two black boxes to start at the same temperature, and then spontaneously, to
have one black box get progressively hotter, and the other to get
progressively colder, over the course of time.

However, the above statements of the second law are false, for one can
falsify them with a single counter-example, namely, two bodies undergoing a
gravitational interaction with each other. Consider, for example, two
identical binary stars which are initially in close proximity, so that their
photospheres are \textquotedblleft kissing\textquotedblright\ each other.
Then there can arise a gravitational instability in which mass is being
transferred from one star to the other. The gravitational virial theorem
leads to the conclusion that the star which is losing mass will be
increasing in temperature, and will become progressively brighter over time,
whereas its companion, which will be gaining mass over time, will be
decreasing in temperature, and will become progressively dimmer over time 
\cite{Lynden-Bell}. This counter-example demonstrates that gravitational
instabilities can lead to thermal instabilities, which is consistent with
the situation shown in Figure 1.

Next, let us consider the second apparent thermodynamic \textquotedblleft
paradox.\textquotedblright\ \textquotedblleft Black holes\textquotedblright\
are \textquotedblleft black,\textquotedblright\ in the sense that they are
perfect absorbers of every kind of particle, including photons at all
frequencies \cite{Sum rule paradox}. Once particles have passed through the
event horizon of a black hole, they can never get out again, at a classical level of
description. For in order for a particle to be able to escape from the black
hole, it would need somehow to acquire an escape velocity which \emph{%
effectively} (in a sense to be defined more precisely below) exceeds the
speed of light at the event horizon.

Since a classical black hole is a perfect absorber, it must also be an ideal blackbody
absorber that could not differ from a blackbody cavity at zero temperature. For
example, a photon, once it has entered through a small aperture of an ideal, cold
blackbody cavity, is irreversibly absorbed by this cavity, never again to be able
to escape to infinity through this aperture (for all practical purposes).
This kind of irreversible behavior of a blackbody cavity at zero temperature is no
different from the irreversible behavior of a classical black hole.

However, in contrast to the case of a classical black hole, a blackbody cavity at a finite
temperature must be able to \emph{emit}, as well as to \emph{absorb} photons 
\cite{Kirchhoff}. Otherwise, the second law of thermodynamics would be
violated. Hawking \cite{Hawking}\ suggested that, like a hot blackbody\ cavity, a
quantum black hole would possess a finite temperature, and therefore must be able to
emit particles, as well as to absorb them. But in order for a black hole to be able
to emit as well as to absorb particles just like the hot blackbody cavity, there would
have to be an effective modification of the no-faster-than-$c$
particle-velocity property of the classical black hole. For only then could a
particle penetrate through the event horizon with an effective escape
velocity which exceeds $c$, and thereby be enabled to escape to infinity as
a free particle.

Since such faster-than-$c$ particle velocities at the event horizon are
fundamentally impossible, classically speaking, it would be impossible for a
classical black hole ever to emit any particles at all. Therefore, a
classical black hole would possess an unusual kind of irreversibility, in which the
black hole could capture particles, but could never release these captured
particles, once they have passed through the event horizon.

However, this kind of irreversibility of a classical black hole would lead to an
apparent violation of the second law of thermodynamics, since the entropy,
for example, the disorder in an ambient photon \textquotedblleft
gas\textquotedblright\ contained in the thermal photons surrounding the black hole,
would be swallowed up by such a classical black hole, and disappear. However, the
mass of a classical black hole would not be increased by the swallowed photons,
since, to an observer at infinity, these photons would have suffered an
infinite gravitational redshift, and would therefore have a zero energy as
they are being swallowed up.\ Hence there would be a decrease of the total
entropy of the entire universe consisting of the unaltered mass of a
classical black hole plus the altered matter and radiation outside of the black hole. In
other words, the total entropy of the universe would have to decrease
steadily over time as the unaltered, classical black hole steadily swallows up the
thermal photons in its vicinity, but leaves no trace whatever of the
disorder that was originally present in the swallowed photon
\textquotedblleft gas.\textquotedblright\ This would contradict the second law of thermodynamics. 

In order to \textquotedblleft save\textquotedblright\ the second law,
Bekenstein \cite{Bekenstein} had to introduce the property of the entropy of
a quantum black hole, which is proportional to the area of the black hole, and to
\textquotedblleft generalize\textquotedblright\ the second law, so that the
entropy of the quantum black hole must be added to the entropy of its surroundings
in order to get the total entropy of the universe. In this way, the total
entropy of the universe could be shown to increase, rather than to decrease,
steadily over time \cite{Bekenstein}\cite{Zurek}.

In contrast to a \emph{classical} black hole, where the event horizon has a zero
width, a \emph{quantum} black hole would possess a \textquotedblleft
fuzzy\textquotedblright\ event horizon with a nonzero width. Such quantum
\textquotedblleft fuzziness\textquotedblright\ necessarily arises from the
uncertainty principle. Bekenstein conjectured that it was the Compton
wavelength that was the relevant length scale for the quantum width of the
event horizon. This length scale arises from a fundamental quantum
uncertainty as to whether a particle of mass $m$ (an electron, say), which
has somehow been localized within a Compton wavelength of the event horizon,
would be captured by the black hole, or not. When this particle is known to have
been localized within the Compton wavelength $h/mc$ of the horizon, its
momentum must have an uncertainty of the order of $\pm mc$, i.e., it will be
fundamentally uncertain whether the particle is moving towards the black hole, or
moving away from the black hole, with a speed on the order of the speed of light $c$%
. Hence it will be uncertain whether a particle of mass $m$ would be
swallowed by the black hole, or would, instead, be able to escape away from the black hole
as a free particle flying off to infinity with an initial speed effectively
exceeding $c$ at the event horizon \cite{emission}. Consequently, there is
one bit of information when the following yes-or-no question is answered:
Did this particle actually get swallowed up by the black hole, or not? This is the
origin of Bekenstein's black hole entropy.

However, the localization of a particle to within a Compton
wavelength of the event horizon of a black hole would also imply the possibility of
pair creation and pair annihilation of particles and anti-particles near the
horizon, e.g., electron-positron pair creation and pair annihilation. For
if the momentum of the particle has an uncertainty of the order of $\pm mc$,
the typical uncertainty in the energy of the particle will be on the order
of $\pm mc^{2}$, which means that there would be enough energy to create
pairs momentarily within the localization layer on the order of a Compton
wavelength in thickness just outside the horizon, which we shall call a
\textquotedblleft Compton layer.\textquotedblright\ Hence it is natural, due
to these uncertainty-principle considerations, to look at pair creation and pair 
annihilation processes in which the net result is that, say, an electron (or a
positron) could be created that could escape to infinity, and
simultaneously, a positron (or an electron) could be captured by the
black hole.

In fact, Hawking has pointed out one such process that could enable the
penetration through the event horizon by a particle, namely, the process of
quantum tunneling, which is similar to the process of the field emission of
electrons from a sharp tip of a charged conductor \cite{Wilczek}. In the case of the
Hawking radiation, it is the presence of the strong tidal gravitational
fields of a black hole at the event horizon, rather than of strong electric fields
at the sharp tip of a charged conductor, that could tear apart virtual,
fluctuating vacuum pairs, which are randomly appearing and disappearing near
the horizon, into free particles that could then escape to infinity. To
quote Hawking, \cite[p.202]{Hawking}:

\begin{quote}
\textquotedblleft Just outside the event horizon [of a black hole] there
will be virtual pairs of particles, one with negative energy and one with
positive energy. The negative particle is in a region which is classically
forbidden but it can \emph{tunnel} [emphasis added] through the event
horizon to the region inside the black hole where the Killing vector which
represents time translations is spacelike. In this region the particle can
exist as a real particle with a timelike momentum vector even though its
energy relative to infinity as measured by the time translation Killing
vector is negative. The other particle of the pair, having a positive
energy, can escape to infinity where it constitutes a part of the thermal
emission described above.\textquotedblright 
\end{quote}

The Hawking tunneling process can be understood in terms of the Feynman
diagram depicted in Figure 2 \cite{unseen-incoming-electron}. 

\begin{figure}
\includegraphics[width=6in]{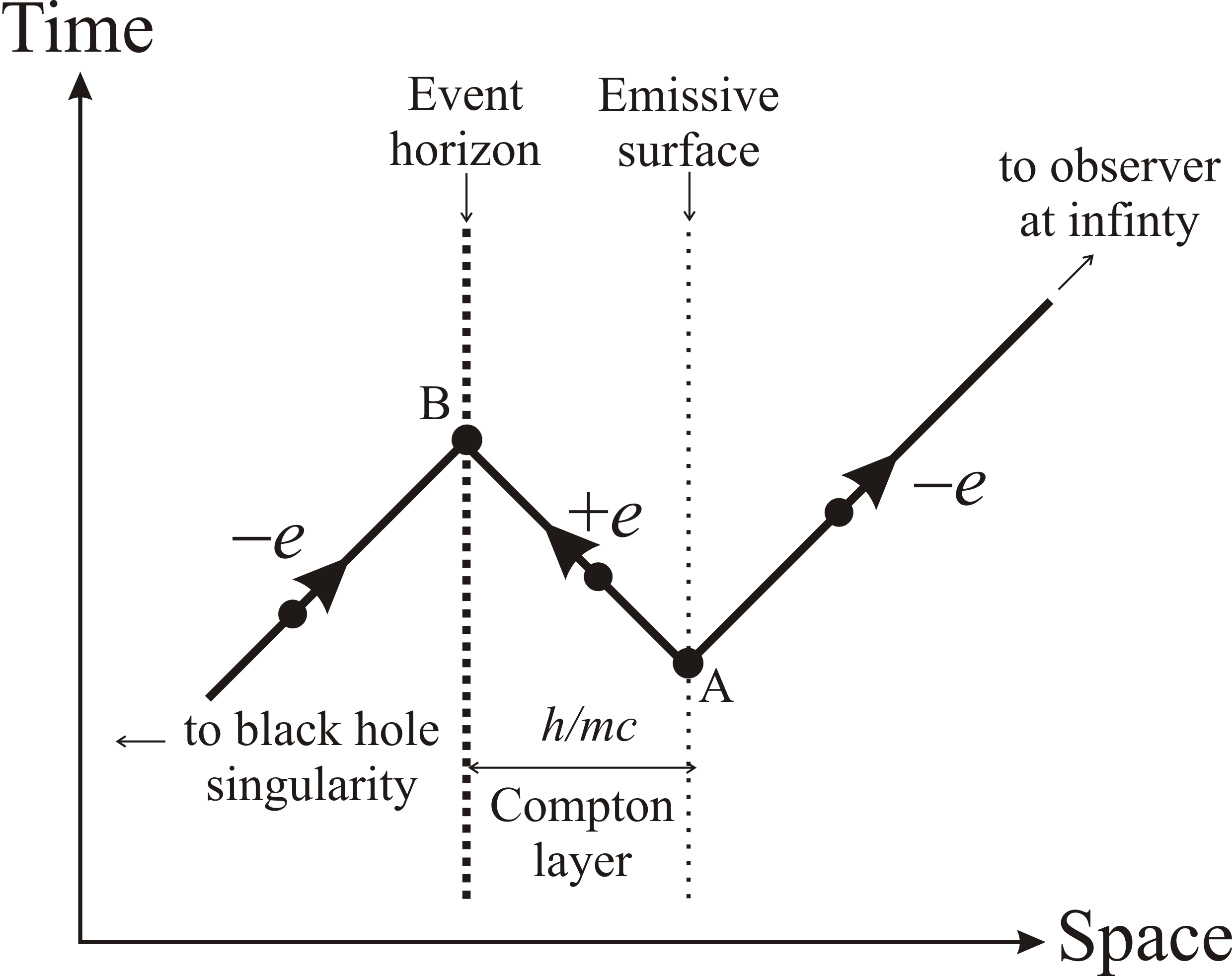}
\caption{Feynman diagram of the Hawking
process for particle creation by a black hole. The heavy, zigzag solid lines
represent the worldlines of electrons ($-e$) and of a positron ($+e$). The
heavy, vertical dashed line on the left represents the event horizon of the
blackhole, and the light, vertical dashed line on the right represents an
effective emissive surface, for the case of electron emission. Points A and
B represent pair creation and pair annihilation, respectively 
\cite{unseen-incoming-electron}. The
\textquotedblleft Compton layer\textquotedblright\ has a thickness on the
order of the Compton wavelength, $h/mc$. The coordinate system being used
here is that of a freely falling observer orbiting around the black hole just
outside of its horizon.}
\end{figure}

Quantum mechanically speaking, the Hawking process could occur when a
particle tunnels, apparently superluminally, through the event horizon,
because its partner could then acquire, through the uncertainty principle,
an \emph{effective} escape velocity which exceeds $c$ at the event horizon.

In particular, thermal photons could then in principle be enabled to escape
to infinity from a black hole. In this way, a quantum black hole could both emit and absorb
particles, just like an ideal blackbody cavity at a finite temperature, and a
violation of the second law of thermodynamics could thereby be avoided.

The idea of a \textquotedblleft superluminal tunneling\textquotedblright\ process
is illustrated by the Feynman diagram in Figure 3 \cite{unseen-incoming-electron}. 
Feynman, in his
\textquotedblleft re-interpretation principle\textquotedblright\ \cite%
{Feynman},\ showed that one can re-interpret a positron going \emph{forwards}
in time as an electron going \emph{backwards} in
time. Hence in Figure 3, instead of a positron propagating forwards in time
from point A to point B as in Figure 2, one can re-interpret this portion of
the Feynman diagram as representing an electron propagating backwards in
time from point B to point A. This can lead to an \emph{effectively}
superluminal speed for the emitted electron.

\begin{figure}
\includegraphics[width=6in]{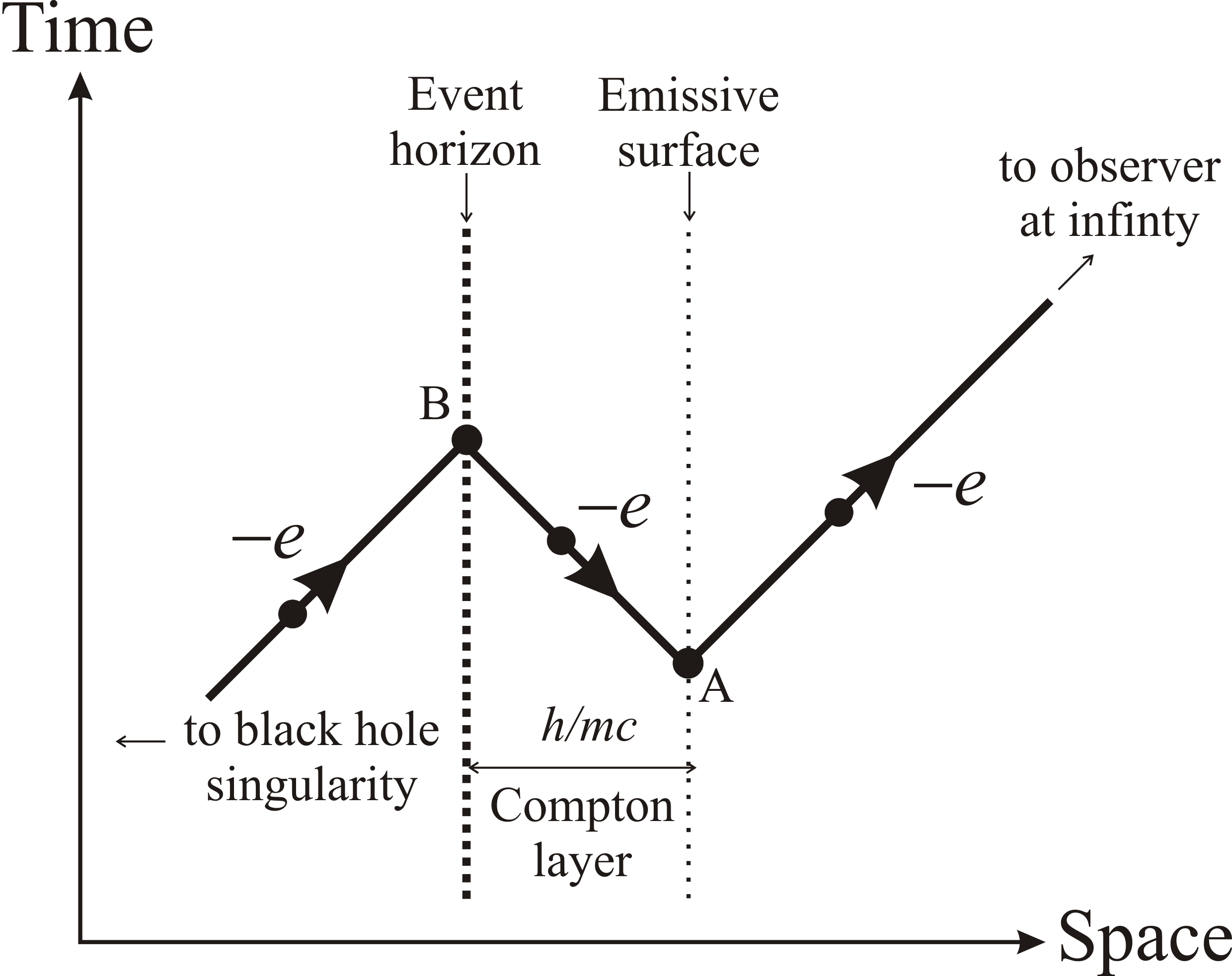}
\caption{Equivalent Feynman diagram of the Hawking process for
particle creation by a black hole. The heavy, zigzag solid line represents
the worldline of a single electron ($-e$), in which the electron is going 
\textit{backwards} in time from B to A, instead of a positron ($+e$) going 
\textit{forwards} in time from A and B as in Figure 2 \cite{unseen-incoming-electron}. The 
resulting electron emission process is effectively superluminal (see text). 
It may be objected that, since the Compton layer is only $h/mc$ thick, 
these superluminal effects will be tiny. However, Hawking \cite%
{Hawking} has pointed out that the Compton wavelength of a photon, 
which has a zero rest mass, is infinite.}
\end{figure}

In order to see this, let us assume for the moment that the electron in
Figure 3 is highly relativistic, i.e., its speed is very close to $c$. By
inspection of Figure 3, it is clear that the electron emitted from point A
would have arrived \emph{earlier} at infinity at a detector
than if it were to have been emitted from point B. Therefore if the observer
at infinity had neglected the finite quantum thickness of the Compton layer
in his calculation of the speed of the electron, and considered only the
case of a classical black hole which could only emit electrons starting at the
zero-width event horizon at point B, then it would have seemed to this
observer that the emitted electron, which started from the event horizon, must have had an effective speed that
exceeded $c$, which would be classically impossible.

But what is impossible classically may sometimes be possible quantum
mechanically. Tunneling through a classically forbidden region of space is
an example. Now Figure 2 is a Feynman diagram representing the Hawking
tunneling process in which a positron tunnels from the \emph{outside} to
the \emph{inside} of a black hole. But Feynman \cite{Feynman} showed that the two
diagrams in Figures 2 and 3 are fully equivalent to each other.
Hence an inverse Hawking tunneling process, in which an electron tunnels
from the \emph{inside} to the \emph{outside} of the black hole, as represented by Figure 3, must be equivalent to the 
process represented by Figure 2.

The phenomenon of \emph{superluminal} tunneling by quantum particles,
including photons, which tunnel through a classically forbidden region of
space with an effective group velocity which exceeds $c$, has been
experimentally observed. The data \cite{Steinberg}\ support Wigner's theory
of the tunneling time \cite{Wigner}, where it was predicted that the time it
takes for a particle to traverse a tunnel barrier is independent of the
thickness of the barrier, for the case of thick, opaque barriers. This
counter-intuitive property of the Wigner time follows from the energy-time
uncertainty principle, which, when applied to a thick, opaque tunnel
barrier, implies a tunneling time that depends only on the energy of the
escaping particle, but does not depend upon the distance traversed by the
particle. Such a superluminal Wigner tunneling time is not forbidden by
relativistic causality, because an effective group velocity for a particle
can, in general, exceed $c$, whereas Sommerfeld's front velocity cannot \cite%
{Wigner}.

Thus the process of quantum tunneling in the case of individual photons has
been experimentally observed to be superluminal, in the sense that the
effective group velocity of the tunneling photon exceeds $c$. Hence it may
be possible for one member of a virtual photon pair produced in a quantum
fluctuation near the event horizon of a black hole to tunnel superluminally from the
outside of the event horizon to the inside, and for its partner to escape to
infinity as a real photon, just like the escaping electron of Figure 2 in
the Hawking tunneling process. The escaping member of the photon pair, which
has been torn apart by the strong gravitational tidal forces near the event
horizon of a black hole, will then appear as if it were a real blackbody photon
within a Planck spectrum to a distant observer, but actually it remains in
an entangled state with respect to the other member of the photon pair that
was swallowed up by the black hole, just as the positron in Figure 2 that was
swallowed up by the black hole will remain in an entangled state with the escaping
electron.

The data \cite{Steinberg}\ agreed with the prediction of the Wigner
tunneling time \cite{Wigner}%
\begin{equation}
\tau =\hbar \left. \frac{d\phi (E)}{dE}\right\vert _{E_{0}}=\hbar \left. 
\frac{d\arg (T(E))}{dE}\right\vert _{E_{0}}  \label{Wigner time}
\end{equation}%
where $\phi (E)=\arg (T(E))$ is the phase of the complex transfer function
for the tunnel barrier, $T(E)$, and where $E$ is the energy of the particle.
The time $\tau $ is to be evaluated at the energy $E_{0}$ of the escaping
particle. The physical meaning of $\tau $ is that it is the time difference
between the time at which the peak of a wavepacket exits a classically
forbidden region of space, and the time at which the peak of wavepacket
enters this region. Wigner introduced this time, which has also been called
\textquotedblleft the group delay,\textquotedblright\ in order to answer the
following question: What is the delay time for the transmission of a
wavepacket through a one-dimensional quantum mechanical tunnel barrier,
whose transfer function $T(E)$ is known?

For thick, opaque tunneling barriers, it can be shown that this time becomes
independent of the thickness of the barrier, and becomes approximately%
\begin{equation}
\tau \simeq \hbar \frac{1}{\Delta E}
\end{equation}%
where, for the case of a rectangular tunnel barrier, $\Delta E$ is the
energy difference between the height of the barrier and the energy of the
incident particle. This result is consistent with the energy-time
uncertainty principle, and is manifestly independent of the width of the
tunnel barrier. For thick, opaque tunnel barriers, therefore, the tunneling
process becomes superluminal.

There have been many theoretical \textquotedblleft tunneling
times\textquotedblright\ which have been suggested, some of which are
superluminal, and others subluminal. Therefore there have been many
controversies concerning which tunneling time is the \textquotedblleft
correct\textquotedblright\ one. Experiments were required to settle these
controversies. In the course of performing these experiments \cite{Steinberg}%
, it has become clear that one must carefully specify the operational
procedure by which the \textquotedblleft tunneling time\textquotedblright\
is actually measured. 

In the case of the black hole, the operational procedure is for an observer outside
of the horizon of the black hole to set up a spectrometer and detector that
can detect whether a thermal photon has escaped the black hole, or not, and, if
it has escaped, to measure the energy of the escaping photon. In this
operational procedure, it is the superluminal Wigner tunneling time given by
(\ref{Wigner time}) that is the relevant tunneling time scale.

However, in the operational method that we used to measure superluminal
photon tunneling times, it was crucial to be able to \textit{compare} the
two arrival times of two twin photons which were produced in a photon pair
production process using parametric down conversion. The two twin photons
traveled through equal distances along the two equal arms of a two-photon
interferometer, i.e., the Hong-Ou-Mandel interferometer, except that one of
the photons had to traverse a tunnel barrier in order to be able to reach
the final beam splitter. This allowed a precise comparison of the arrival
times of the two twins, one of which had tunneled through a barrier, and the
other which had not. This comparison was accomplished using coincidence
detection by means of two Geiger counters placed at the two output ports of
the final beam splitter of the Hong-Ou-Mandel interferometer. Our results confirmed Wigner's superluminal tunneling time.

The equivalent operational method in Figure 3 would be to compare the
arrival times of a photon emitted from point A with one that is emitted from
point B. However, since point B lies exactly on the event horizon, it is
questionable whether any emission could ever occur, since no particles,
including photons, could ever escape from the horizon \cite{white hole}. Hence it remains an
open question whether or not the superluminal Wigner tunneling time is
actually applicable to Hawking radiation from a black hole, operationally speaking. 

Nevertheless, one can appeal solely to Figure 2, without any appeal to
Figure 3, in order to justify\ the existence of Hawking radiation. Feynman's re-interpretation principle can be applied to an electron
moving backwards in time, i.e, superluminally, so that the electron can be
replaced with a positron moving forwards in time, i.e., without any
superluminality being invoked. One could thus evade the question of
whether or not the superluminality of tunneling actually occurs in black
hole emission. However, this evasion would overlook the equivalence of the two
Feynman diagrams shown in Figures 2 and 3.

The author would like to thank Jacob Bekenstein, Sam Braunstein, Bill Unruh and Bob Wald for
very helpful discussions.


\begin{thebibliography}{99}
\bibitem{fluctuation} Such fluctuations must necessarily arise due to the
discrete nature of photons which are contained in blackbody radiation.
\bibitem{Sum rule paradox} Another \textquotedblleft
paradox,\textquotedblright\ which will not be addressed here, involves the
infinite-frequency sum rule for the integral of the cross section $\sigma
_{abs}(\omega )$ for the absorption of photons by a classical black hole
over all frequencies $\omega $, i.e.,%
\begin{equation}
\int\limits_{0}^{\infty }\sigma _{abs}(\omega )d\omega =\infty 
\end{equation}%
which seems to be inconsistent with the Kramers-Kronig relations and the
principle of causality from which this rule is derived.
\bibitem{emission} Note that this uncertainty-principle argument therefore also
implies that a quantum black hole must be able to $emit$, as well as to $absorb$, a
particle of mass $m$, when this particle is localized within a
\textquotedblleft fuzzy\textquotedblright\ event horizon with a quantum
width of a Compton wavelength $h/mc$, and thereby also avoids a 
violation of the second law of thermodynamics.
\bibitem{unseen-incoming-electron} It may be objected that the portion of the Feynman 
diagram to the left of point B in Figure 2 (and also in Figure 3) is physically meaningless, 
since the observer outside of the event horizon cannot see this portion of the incoming 
electron's trajectory, which lies in a physically inaccessible portion of spacetime. All 
that the outside observer can see is the sudden disappearance of the positron at point 
B in Figure 2 (or, by a similar argument, the sudden appearance of the electron at point 
B in Figure 3), and nothing more. However, the charge of the positron cannot simply 
disappear into nothingness at point B in Figure 2 (nor can the charge of the electron 
suddenly appear out of nothingness at point B in Figure 3). Otherwise, the principle of 
charge conservation would be violated. Thus charge conservation necessitates 
the annihilation of the positron's positive charge by an electron's negative charge from 
an unseen incoming electron which originates from beyond the event horizon in Figure 2. 
Moreover, the principle of momentum conservation necessitates the outgoing electron's 
momentum, as it emerges from point A in Figure 2, to be exactly equal to the unseen 
incoming electron's momentum as it approaches point B from the left. Hence the outside 
observer would infer the existence of the unseen incoming portions of the Feynman 
diagrams which would extend beyond point B into the interior of the black hole in both 
Figures 2 and 3.
\bibitem{white hole} However, note that emission from point B of a photon, as well as emission from point A of a photon, is possible from a $white$ hole, in contrast to the case of a $black$ hole.
\end{thebibliography}
\end{document}